\documentclass[
 reprint, superscriptaddress, amsmath, amssymb, prl
 ]{revtex4-2}

\usepackage{graphicx}
\usepackage{dcolumn}
\usepackage{bm}
\usepackage{xcolor}
\usepackage[normalem]{ulem}
\definecolor{prlblue}{RGB}{46, 48, 146}
\usepackage[colorlinks=true,citecolor=prlblue,urlcolor=prlblue,linkcolor=black]{hyperref}

\begin{document}
\preprint{Preprint -- Not for Distribution}

\title{A single-crystal alkali antimonide photocathode: high efficiency in the ultra-thin limit}

\author{C. T. Parzyck}
  \altaffiliation{C.T.P and A.G. contributed equally to this work.}
  \affiliation{Laboratory of Atomic and Solid State Physics, Department of Physics, Cornell University, Ithaca, NY 14853, USA}
\author{A. Galdi}
  \altaffiliation{C.T.P and A.G. contributed equally to this work.}
  \affiliation{Cornell Laboratory for Accelerator-Based Sciences and Education, Cornell University, Ithaca, NY 14853, USA}
\author{J. K. Nangoi}
  \affiliation{Laboratory of Atomic and Solid State Physics, Department of Physics, Cornell University, Ithaca, NY 14853, USA}
\author{W. J. I. DeBenedetti}
  \affiliation{Department of Chemistry and Chemical Biology, Cornell University, Ithaca, NY 14853, USA}
\author{J. Balajka}
  \affiliation{Department of Chemistry and Chemical Biology, Cornell University, Ithaca, NY 14853, USA}
\author{B. D. Faeth}
  \affiliation{Platform for the Accelerated Realization, Analysis, and Discovery of Interface Materials (PARADIM), Cornell University, Ithaca, NY 14853, USA}
\author{H. Paik}
  \affiliation{Platform for the Accelerated Realization, Analysis, and Discovery of Interface Materials (PARADIM), Cornell University, Ithaca, NY 14853, USA}
\author{C. Hu}
  \affiliation{Laboratory of Atomic and Solid State Physics, Department of Physics, Cornell University, Ithaca, NY 14853, USA}

\author{T. A. Arias}
  \affiliation{Laboratory of Atomic and Solid State Physics, Department of Physics, Cornell University, Ithaca, NY 14853, USA}
\author{M. A. Hines}
  \affiliation{Department of Chemistry and Chemical Biology, Cornell University, Ithaca, NY 14853, USA}
\author{D. G. Schlom}
  \affiliation{Department of Materials Science and Engineering, Cornell University, Ithaca, NY 14853, USA}
  \affiliation{Kavli Institute at Cornell for Nanoscale Science, Cornell University, Ithaca, NY 14853, USA}
  \affiliation{Leibniz-Institut f{\"u}r Kristallz{\"u}chtung, Max-Born-Stra{\ss}e 2, 12489 Berlin, Germany}
\author{K. M. Shen}
 \affiliation{Laboratory of Atomic and Solid State Physics, Department of Physics, Cornell University, Ithaca, NY 14853, USA}
  \affiliation{Kavli Institute at Cornell for Nanoscale Science, Cornell University, Ithaca, NY 14853, USA}
\author{J. M. Maxson}
  \email[Corresponding Author: ]{jmm586@cornell.edu}
  \affiliation{Cornell Laboratory for Accelerator-Based Sciences and Education, Cornell University, Ithaca, NY 14853, USA}

\date{\today}

\begin{abstract}
The properties of photoemission electron sources determine the ultimate performance of a wide class of electron accelerators and photon detectors. To date, all high-efficiency visible-light photocathode materials are either polycrystalline or exhibit intrinsic surface disorder, both of which limit emitted electron beam brightness. In this letter we demonstrate the synthesis of epitaxial thin films of Cs$_3$Sb on 3C-SiC (001) using molecular-beam epitaxy. Films as thin as 4 nm have quantum efficiencies exceeding 2\% at 532 nm. We also find that epitaxial films have an order of magnitude larger quantum efficiency at 650 nm than comparable polycrystalline films on Si. Additionally, these films permit angle-resolved photoemission spectroscopy measurements of the electronic structure, which are found to be in good agreement with theory.  Epitaxial films open the door to dramatic brightness enhancements via increased efficiency near threshold,  reduced surface disorder, and the possibility of engineering new photoemission functionality at the level of single atomic layers. 
\end{abstract}

\maketitle

Dense and coherent electron beams generated via photoemission have enabled many revolutionary tools in the physical sciences. These include x-ray free electron lasers with $10^9$ times higher peak x-ray brightness than previous sources \cite{LCLS, euxfel}, ultrafast electron microscopes that achieve femtosecond temporal and atomic structural resolution \cite{4dem, dwayne},  hadron collider luminosity-enhancing systems based on electron beams \cite{cec}, and next-generation electron linear colliders \cite{ccubed, ilc}, which will probe physics beyond the standard model. Electron beam brightness, defined as the density of the beam in position-momentum phase space, is a central figure of merit for each of these applications \cite{Musumeci}. According to Liouville's theorem, the brightness of any electron beam in a linear accelerator can be no greater than at its source \cite{BazarovMax,Filippetto}. Thus, the photoemission performance of photocathodes can determine the ultimate performance of accelerators ranging from meters to kilometers in length \cite{Pierce}. Photocathodes are also the critical enabling technology for many photon detectors, ranging from photomultipliers \cite{Wright_2017}, image intensifiers and ultrafast time-resolved streak cameras \cite{streakcamera}, to detectors for high energy physics applications \cite{Lyashenko_2020}.

For intense beam generation or sensitive detection applications, many metallic photocathodes are infeasible choices due to their low quantum efficiencies (QE)---the number of emitted electrons per incident photon. For metals, QEs generally range from $10^{-3}$ to $10^{-5}$ with ultraviolet illumination \cite{sommer}. In contrast, the highest quantum efficiency materials, semiconductors containing or coated with alkali metals, have peak quantum efficiencies exceeding 1\% in the the visible, with response extending into the near-infrared \cite{sommer}.  In particular, the spectral response of alkali antimonides extends near the lasing wavelength of common high-bandwidth gain media and therefore can eliminate lossy wavelength conversion stages in both accelerator and detector applications \cite{Musumeci2010,Cultrera2016,tradeoff}.  Additionally, by limiting the required light intensity for a given photoelectron density, high quantum efficiency can mitigate multiphoton photoemission (MPPE).  Since MPPE causes a dramatic increase in the momentum spread of photoelectrons, its suppression can dramatically improve beam brightness \cite{Jai}.

Owing to momentum conservation in the photoemission process, the transverse momentum spread of photoelectrons from clean, ordered, crystalline surfaces can be reduced by exploiting the material's band structure \cite{effectivemass}. For single-crystal metal photocathodes this has been theoretically modeled and experimentally verified \cite{Karkare_Ag, Karkare_Cu}. Record low momentum spread has been achieved in Cu (100) near the photoemission threshold; however, the efficiency ($\sim10^{-8}$) is too low for the beam current requirements of moderate to high-intensity applications \cite{Karkare_Cu}. It is therefore desirable to synthesize maximally-efficient cathodes as single-crystal films, where the smoothness, homogeneity, and termination can be controlled.

To date, all high-efficiency visible light photocathodes are either grown as (often defect-rich) polycrystals, or produced by activating a semiconductor surface (e.g., GaAs) with an alkali metal, commonly cesium, resulting in an intrinsically disordered emitter surface \cite{YAMADA, Fukuzoe, Biswas}. Owing to their alkali metal content, these photocathodes are extremely sensitive to vacuum contamination and cannot be characterized \textit{ex situ}; a single Langmuir of oxygen is sufficient to alter their efficiency and surface chemistry \cite{Bates1981, DANIELSON, Soriano1993, Hines}. While great progress has been made in the growth of alkali antimonide films using evaporative and sputtering techniques \cite{Feng2017,Ruiz-Oses2014a,Schubert2016,Gaowei2017,Ding2017,Xie}, smooth, epitaxial, single--phase samples remain elusive.  This presents a barrier not only to the reduction of physical \cite{Feng-rough, Galdi-roughness} and surface potential \cite{roughnessGaAs} roughness, but also to the engineering of these unique low-work function semiconducting materials.  

In this Letter, we demonstrate the first epitaxial growth of thin films of the high-efficiency photocathode material Cs$_3$Sb via molecular-beam epitaxy.  The resulting films were characterized using a variety of \textit{in situ} diagnostics including reflection high-energy electron diffraction (RHEED), x-ray and ultraviolet photoemission spectroscopy (XPS, UPS), and angle-resolved photoemission spectroscopy (ARPES).  Structural and spectroscopic measurements of our films agree well with both previous studies of polycrystalline samples and density functional theory (DFT) calculations of the electronic structure. 

\begin{figure}
  \resizebox{1\columnwidth}{!}{  
  \includegraphics{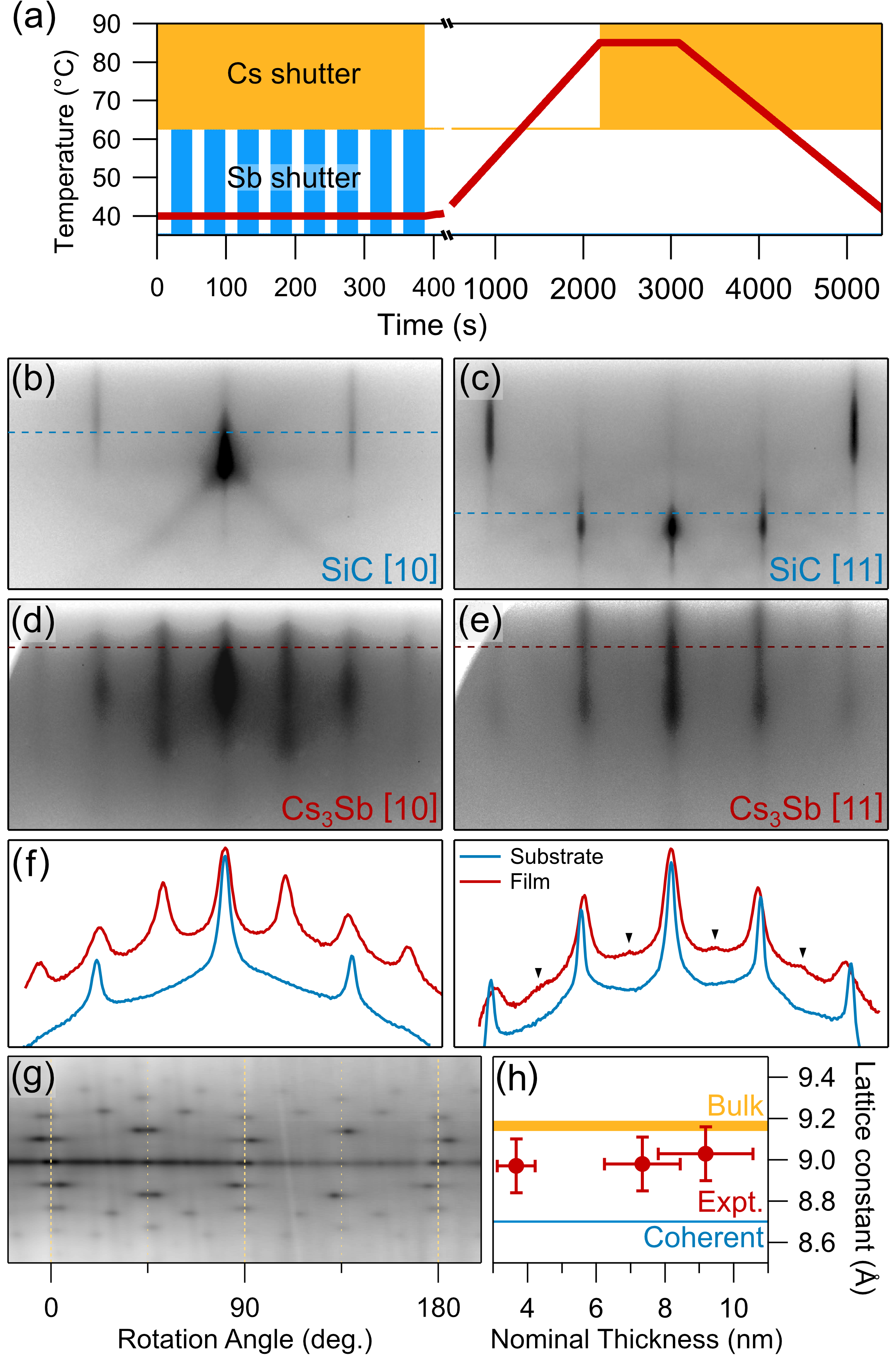}}
  \caption{\label{growth} Growth and structural characterization. (a) Substrate temperature (red line) and shutter states (filled areas: shutter open) during the growth/recrystallization process of 2 u.c. of (001)-oriented Cs$_3$Sb. RHEED images of an annealed SiC substrate (b-c) and a 10 u.c. Cs$_3$Sb film (d-e) along corresponding azimuths. (f) Line profiles of the substrate (blue) and film (red) along dashed lines in the above RHEED images (offset for clarity); half-order peaks are marked with black triangles. (g) RHEED line profile intensity of a 10 u.c. film as a function of the azimuthal angle ($0^{\circ}=[10], 45^{\circ}=[11]$). (h) In-plane lattice constant of three samples, measured using RHEED, grown under nominally identical conditions.  Orange and blue lines mark the bulk Cs$_3$Sb lattice parameter and  $2\times a_{SiC}$, respectively. Panels (b)-(g) use a logarithmic intensity scale.}
\end{figure}

Cs$_3$Sb has a photoemission threshold $>600$~nm, QE exceeding 1\% at 530~nm, and crystallizes exclusively in a cubic structure with lattice constant, $a_{bulk}$, in the range of 9.14 to 9.19~\AA~\cite{sommer,Jack1957,Gnutzmann1961,McCarroll1965,Barois1989,Sangster1997, Robbie1973}. Earlier work identified a $F_{d\bar{3}m}$ space group (NaTl-like)\cite{Jack1957, McCarroll1965}, while more recent studies favor an ordered $F_{m\bar{3}m}$ Cu$_3$Al structure \cite{Gnutzmann1961, Robbie1973}. The large cubic lattice constant presents a challenge in the selection of a suitable substrate, however, recent studies have identified 3C-SiC as a promising candidate due to its lattice parameter, $a_{SiC}=4.36$ \AA, nearly matching $a_{bulk}/2 \approx 4.59$ ($\sim5\%$ strain), and the stability of its (001) surface \cite{Galdi-roughness, GaldiIPAC}.

Epitaxial thin films were grown on cubic 3C-SiC~(001) substrates (epitaxial film on Si, MTI corporation) via MBE, using a sequence of shuttered growth of two unit-cells (u.c.) at a substrate temperature of 40~$^\circ$C and a recrystallization step at 85~$^\circ$C. This process is described schematically in Fig.~\ref{growth}~(a). During the growth phase, Sb is provided in eight doses of $1/4$ nominal monolayers per SiC unit-cell, separated by a 20~s pause, while Cs is provided continuously. The sample is then heated to the recrystallization temperature in vacuum, annealed in Cs flux for 15~min, and cooled in Cs flux back to the growth temperature. Each cycle produces nominally 2 unit cells of Cs$_3$Sb, within the $\pm 15\%$ uncertainty of the quartz-crystal microbalance measurement of the Sb flux.  This growth process departs from traditional co-deposition procedures in favor of a solid-phase epitaxy approach as we have observed that films grown by codeposition at a single temperature are either crystalline or efficient photoemitters, but not both simultaneously.  Films codeposited at low temperatures ($<80$~$^{\circ}$C) appear stoichiometric and have high QE, but lack ordered surfaces.  Conversely, films codeposited at higher temperatures (90-150~$^{\circ}$C) form an ordered crystalline phase, but are generally Cs deficient (measured by XPS) and have low QE ($<1\times10^{-4}$ at 532 nm).  A low deposition temperature is necessary to reduce the Cs desorption rate and produce the correct stoichiometry; then the anneal is required to improve crystallinity and order the domains to produce an epitaxial film.  

RHEED patterns of a SiC substrate and a 10 u.c. (nominally 9.2-9.5 nm) thick film are displayed in Fig.~\ref{growth}~(b-e). Comparison of the film diffraction patterns with the reciprocal lattice of Cs$_3$Sb confirm the cube-on-cube epitaxial relationship between film and substrate: Cs$_3$Sb $\langle 100 \rangle$~//~SiC  $\langle 100 \rangle$. The in-plane ordering of the the film is confirmed by a RHEED line profile intensity map presented in Fig.~\ref{growth}~(g). This map shows both a $90^\circ$ period, consistent with cubic symmetry, and no repetition of the Bragg peaks within a $45^\circ$ window, precluding the presence of rotationally misaligned macroscopic domains. From the RHEED diffraction streak spacing, the in-plane lattice constant for the thinnest (4 u.c.) film is found to be $a_{film}=8.97\pm0.13$ \AA~, indicating partial relaxation ($2\times a_{SiC}<a_{film}<a_{bulk}$) at a thickness of only 3.8 nm.  In Fig.~\ref{growth}~(e), weak half-order streaks are present in the RHEED pattern along the [11] azimuth, evidenced by the line profiles in (f). These correspond to forbidden odd-index reflections in either the $F_{m\bar{3}m}$ or the $F_{d\bar{3}m}$ Cs$_3$Sb structure; their presence may signal the presence of lattice distortions (either intrinsic or induced by defects or vacancies) or a surface reconstruction.   

\begin{figure}
  \resizebox{1\columnwidth}{!}{  
  \includegraphics{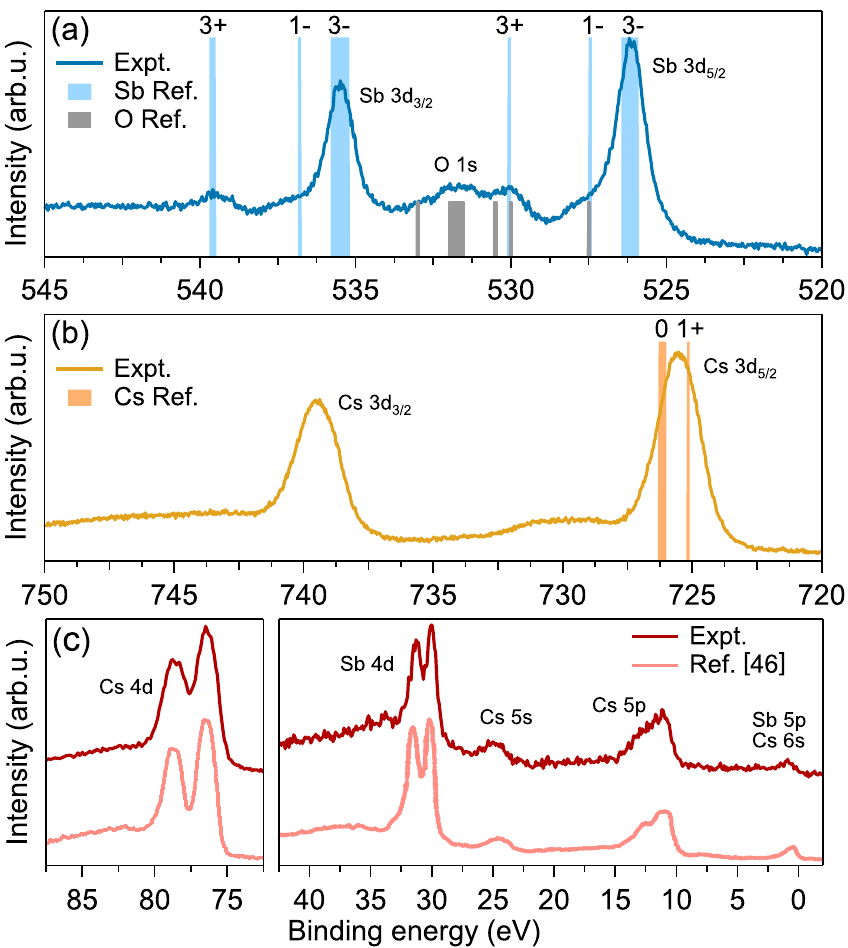}}
  \caption{\label{PE} XPS Spectra of a 7.4 nm thick epitaxial Cs$_3$Sb film. (a) Sb $3d$ and O $1s$ regions. Cyan bars indicate binding energy references for different Sb species; gray bars denote the O $1s$ references for different oxides. (b) Cs $3d$ region of the same sample. Orange bars indicate the literature references for the Cs oxidation states. (c) Valence band spectra including Cs $4d$, $5s$, $6s$ and Sb $4d$ and $5p$ regions.  Data from a high efficiency polycrystalline Cs$_3$Sb film (reproduced from \cite{Bates1980}) are provided for reference.}
\end{figure}

To confirm the formation of the Cs$_3$Sb phase, \textit{in situ} XPS measurements were performed and typical spectra are reported in Fig.~\ref{PE}. The dominant peaks in the Sb region correspond to the 3- valence, as expected for Cs$_3$Sb, while weak features corresponding to more oxidized species (1-, 3+) are also visible at higher binding energy. Between the Sb $3d_{3/2}$ and $3d_{5/2}$ peaks, weak peaks assigned to O $1s$ states of various Cs and Sb oxides are also visible; typical of Cs$_3$Sb that has been exposed to trace amounts of oxygen \cite{Hines, Bates1981, Soriano1993}. We attribute the surface oxidation to reaction with residual gasses in the UHV transfer system.  The Cs $3d_{5/2}$ peak, Fig.~\ref{PE}(b), is shifted to slightly higher binding energy than the 1+ reference, attributed to band bending induced by surface oxidation \cite{Hines}, and is accompanied by a strong plasmon peak---typically correlated with high quantum efficiency \cite{Bates1980, MARTINI2015}.  Finally, valence band measurements, Fig.~\ref{PE}(c), agree with data on polycrystalline films reported in \cite{Bates1980}. In particular, the Cs $5s$ to Cs $5p$ peak-area ratios closely match those reported by Bates et al. for high QE Cs$_3$Sb (attributed to minimal excess metallic Cs on the surface).  We find area ratios ranging from 0.23 to 0.30 (or 0.15 - 0.21 if the Sb $4d$ tail contributions are subtracted) for samples with QE ranging from 2.1 to 3.3\%.

\begin{figure}
  \resizebox{1\columnwidth}{!}{  
\includegraphics{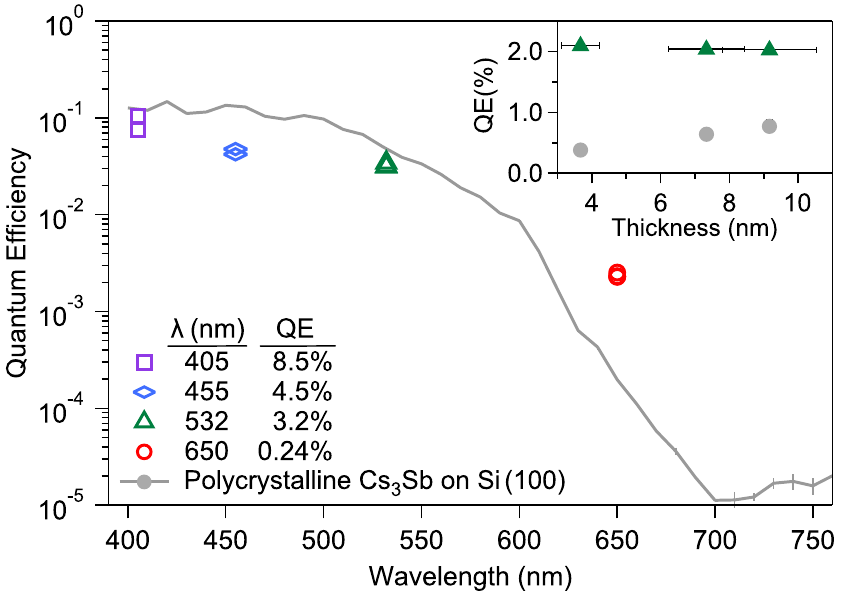}}
\caption{\label{QE} Quantum efficiency, as a function of wavelength, for one sample, measured at two locations on the surface. The QE of a representative codeposited sample grown on Si~(100) is included for comparison. The inset shows the QE, at 532 nm, of 3 samples of varying thickness grown under the same nominal conditions (triangles) and the QE of the codeposited sample on Si whose spectral response is reported in the main panel (dots).} 
\end{figure}

Following growth, the QE of the samples was measured in an adjoining chamber using a biased collection coil. The QE of the $\sim10$~nm (10 u.c.) sample of Fig.~\ref{growth}~(d-e), is shown in Fig. \ref{QE}, demonstrating that its quantum efficiency exceeds 3\% below 532~nm, and is significantly enhanced at 650~nm (0.24\%) compared to typical polycrystalline Cs$_3$Sb on silicon. In the inset, we report the QE (at 532~nm) of a sequentially grown thickness series; the QE exceeds 2\% for the 4 u.c. sample and is not strongly affected by increasing thickness up to 9~nm. It is unlikely that the measured photocurrent includes electrons excited from the substrate since silicon carbide is a 2.3~eV indirect bandgap semiconductor where direct optical transitions require photon energies exceeding 4~eV \cite{Zhao_2000}, well outside the measurement range.  The measured thickness dependence departs from that of previous polycrystalline samples and represents an interesting avenue for future research.    

Thickness is a critical parameter in determining a photocathode's response time---the resulting electron pulse length following illumination by a delta-function optical impulse. For semiconductor photocathodes, this is governed by the transit time of excited electrons to the surface \cite{karkareultra} and is naturally restricted in thin samples. In our case, the electron transit time is expected to be of order $\sim 10$ fs, based on our DFT-based \textit{ab initio} photoemission calculations. This suggests that these thin, epitaxial cathodes are excellent candidates for time-resolved electron beam instruments such as ultrafast microscopes \cite{Hassan_2017}, or high time-resolution streak cameras \cite{streakcamera}.
\begin{figure*}
  \resizebox{2\columnwidth}{!}{ 
  \includegraphics{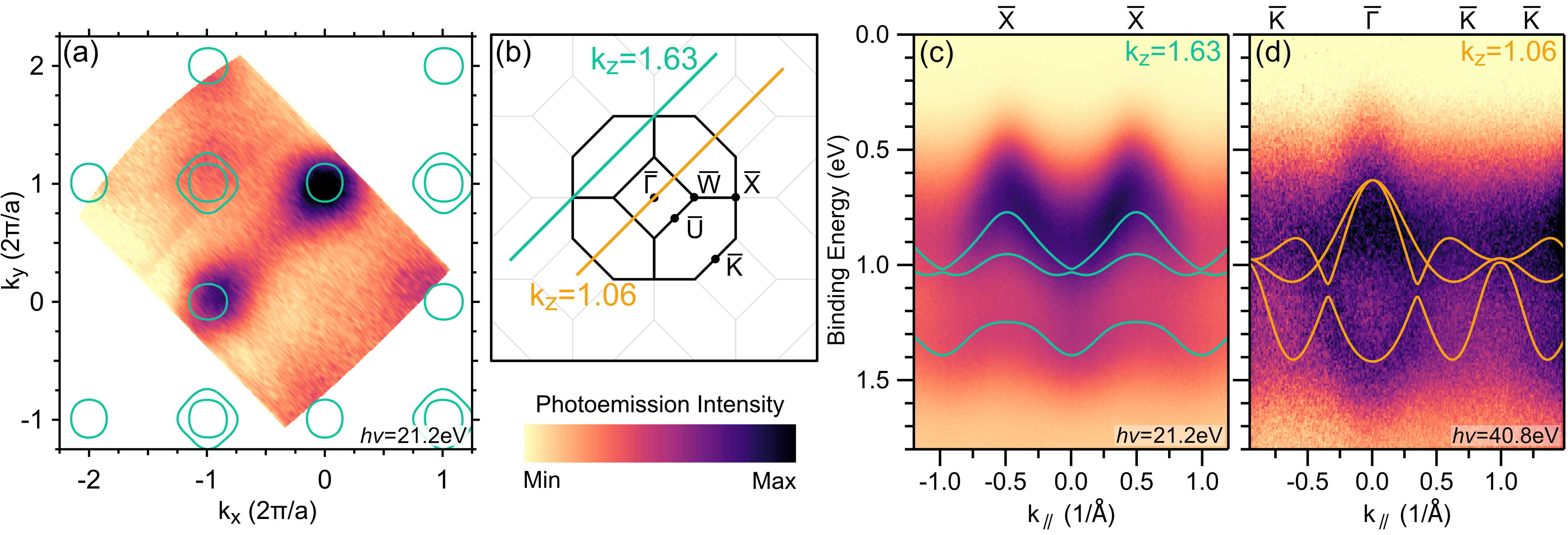}}
  \caption{\label{ARPES} \textit{In situ} ARPES measurements of Cs$_3$Sb thin films. (a) Constant-energy intensity map integrated over a 50 meV window about $E-E_f=-0.55$~eV.  Overlay (green) shows band positions at $E-E_{VBM}=-0.19$~eV calculated by DFT. (b) Schematic of the projected FCC Brillouin zone showing locations of the cuts shown in the following panels. (c) ARPES spectrum passing through the edge of the projected zone at a nominal out-of-plane momentum of $k_z=1.63$~r.l.u.. Corresponding calculated band positions are shown in green. (c) Spectrum taken through the zone center at $k_z=1.03$ r.l.u with calculated bands overlaid in orange.  In panels (c-d) the DFT bands have been offset in energy by -0.63 eV.}
\end{figure*}

There is significant evidence that the electronic band structure, in concert with many-body effects, determines the momentum distribution of photoemitted electrons from atomically ordered surfaces, even when the photon energy is very near the photocathode workfunction \cite{Karkare_Cu, nangoi2020ab, effectivemass,karkare_single_crystal,Karkare_Ag}---where brightness is often largest \cite{tradeoff, Musumeci}. Nonetheless, the minimum mean transverse energy measured in polycrystalline Cs$_3$Sb, $\sim 20$ meV at 90 K \cite{LucaCold}, is a factor of 2.6 larger than the free-electron thermal limit estimate \cite{feng2015}, and an order of magnitude larger than the minimum predicted by our DFT-based ab initio photoemission theory \cite{Nangoi2018}. Fully utilizing the electron band structure and maximizing photoemission brightness requires an atomically ordered surface \cite{karkare_single_crystal, Karkare_Ag, Gevorkian}, which has not been possible in alkali antimonides until now.

A major advantage of the epitaxial growth method presented here is that the resulting Cs$_3$Sb films possess sufficiently ordered surfaces such that the band structure is observed in the photoemitted electrons, and can be measured using ARPES for comparison with theoretical calculations. Measurements were performed at room temperature in a background pressure of $P<1\times 10^{-10}$~torr using a helium plasma discharge lamp and the results are summarized in Fig.~\ref{ARPES}. Due to the highly three dimensional nature of the band structure, determination of the out-of-plane momentum, $k_z$, is required to make quantitative comparisons to DFT calculations.  We estimate the inner potential to be $V_0 = \phi + W$ \cite{Damascelli2004} using the bandwidth $W\approx 1.2$ eV estimated from UPS spectra and the work function $\phi \approx 1.7$ eV measured from polycrystalline samples.  This yields estimates of $k_z\approx 2.4$~\AA$^{-1}$ and $k_z\approx 3.3$~\AA$^{-1}$ for He-I ($h\nu=21.2$~eV) and He-II ($h\nu=40.8$~eV), respectively.  To assign the absolute momenta to positions in the Brillouin zone, in the presence of uncertainty in the out-of-plane lattice constant, $c_{film}$, comparison of the ARPES spectra and DFT calculations is performed across several independent measurements (\textit{c.f.} supplemental materials \cite{supplemental}).  This gives a best fit with $k_z = 1.63$~r.l.u. (1 r.l.u = $2\pi/c_{film}$) for He-I and $k_z = 1.06$~r.l.u. for He-II along with an overall shift of the Fermi level by 0.63 eV (attributable to pinning of $E_f$ in the gap)---consistent with an inner potential of $V_0=3.2$~eV and lattice constant of $c_{film}=9.5$~\AA.

Figure~\ref{ARPES}(a) shows a photoemission intensity map taken with He-I light at $E-E_f=0.550$~eV, just cutting through the tops of the valence bands.  The measured intensity pattern is consistent with a FCC Brillouin zone with $a_{film}=9.0$~\AA.  An overlay showing the calculated tops of the valence bands (at the same $k_z$, and at an energy of $E-E_{VBM}=-0.190$~eV) agrees well with identified band positions.  Figure~\ref{ARPES}(c) shows a cut along the \smash{$\bar{X}-\bar{X}$} direction taken with He-I light and Fig.~\ref{ARPES}(d) a cut along the \smash{$\bar{K}-\bar{\Gamma}-\bar{K}$} line taken with He-II. In the \smash{$\bar{X}-\bar{X}$} cut we observe good agreement between the predicted and observed bands, with one brighter set between 0.5 and 1.0 eV of binding energy separated from a heavier and less intense band between 1.2 and 1.5 eV.  In the \smash{$\bar{K}-\bar{\Gamma}-\bar{K}$} cut we again observe a reasonable match between the predicted and observed dispersions---in particular the hole-like band at \smash{$\bar{\Gamma}$}, at a $k_z$ of 1.06 r.l.u., lies nominally at the $X$ point of the FCC zone and constitutes the valence band maximum.

While there is good qualitative agreement with our calculations, we observe a roughly 10-20\% larger bandwidth in the ARPES spectra than is predicted by DFT.  Preliminary calculations indicate that the Sb $3p$ / Cs $6s$ bandwidth is influenced by the presence of both distortions from the ideal $F_{m\bar{3}m}$ structure and by biaxial compressive strain.  The effects are found to be in competition, with distortions decreasing the bandwidth and strain increasing it, so we cannot disentangle their individual effects at present.  This motivates further \textit{in situ} structural characterization of this epitaxial phase, such as x-ray diffraction measurements \cite{Ruiz-Oses2014a,Schubert2016,Gaowei2017,Ding2017} and a more detailed study of strain effects on the photoemission processes of Cs$_3$Sb.

In conclusion, we have used solid-phase epitaxy to grow (001)-oriented Cs$_3$Sb films on 3C-SiC (001) substrates. These films display unusually high efficiency at thicknesses as low as 4 u.c. ($\sim 3.8$~nm). Thanks to the ordered nature of the sample surface, we have obtained the first ARPES measurements of the band structure, allowing direct comparison with DFT.  As an indicator of what may be possible in the future, epitaxial control in other classes of semiconducting photocathodes has yielded remarkable new capabilities. Epitaxial strain induced in a GaAs/GaAsP superlattices lifts a band degeneracy at the valence band maximum and significantly enhances the spin polarization of the emitted electrons \cite{Maruyama_2004}. An even more intricate epitaxial layered structure enables an interferometric enhancement of absorption in GaAs, improving quantum efficiency by nearly an order of magnitude \cite{Liu_2016}.   Control of the surface dipole moment in epitaxial GaN, an ultraviolet semiconductor photocathode, yields a negative electron affinity surface without cesium, greatly enhancing its intrinsic quantum efficiency and ease of use \cite{Marini_2018}. This work opens the door to similar engineering at the level of single atomic layers in the alkali antimonides: controlled doping, strain, termination, surface dipole moment, and heterostructuring are prime targets of future work for enhanced brightness in these remarkable photoemissive compounds. \newline

The data that support the findings of this study are available within the paper and supplementary material \cite{supplemental}. Additional data related to the growth and structural characterization are available at (DOI pending publication). Any additional data connected to the study are available from the corresponding author upon reasonable request. \newline

This work was supported by the U.S. National Science Foundation Grant PHY-1549132, the Center for Bright Beams, and by the National Science Foundation (Platform for the Accelerated Realization, Analysis, and Discovery of Interface Materials (PARADIM)) under Cooperative Agreement No. DMR-2039380. This work made use of the Cornell Center for Materials Research Shared Facilities which are supported through the NSF MRSEC program (DMR-1719875).

\nocite{*}  
\bibliography{Cs3Sb_arXiv_Draft}
\end{document}